\documentclass[12pt,preprint]{aastex}
\usepackage{url}
\usepackage{natbib}
\usepackage{aas_macros}
\usepackage{hyperref}
\usepackage[usenames, dvipsnames]{color}
\shorttitle{Application of the Synoptic Magnetograms to  Global Solar Activity Forecast}
\shortauthors{Kitiashvili, I.N.}

\title{Application of Synoptic Magnetograms to Global Solar Activity Forecast}
\author{I.~N. Kitiashvili$^{1,2}$}
\affil{$^1$NASA Ames Research Center, Moffett Field, Mountain View, CA 94035, USA}
\altaffiltext{1}{e-mail: irina.n.kitiashvili@nasa.gov}
\affil{$^2$Bay Area Research Enviromental Institute, Moffett Field, Mountain View, CA 94035, USA}
\begin{document}
\begin{abstract}
Forecasting solar activity is a long-standing problem, mainly due to limitations of available observational data and model inaccuracies. Synoptic magnetograms, available for the last four solar cycles, provide us with knowledge about the evolution of magnetic fields on the solar surface and present important information for forecasting future solar activity. In this work, poloidal and toroidal magnetic field components derived from synoptic magnetograms are assimilated, using the Ensemble Kalman Filter method, into a mean-field dynamo model based on Parker's migratory dynamo theory complemented by magnetic helicity conservation. Tests performed by predicting previous solar cycles 23 and 24 show good potential for this approach for prediction of the upcoming solar cycle and support future developments of this methodology. It was found that the predicted toroidal field is in good agreement with observations for almost the entire following solar cycle. However, poloidal field predictions agree with observations only for the first 2 -- 3 years of the predicted cycle. The results, based on the synoptic magnetograms obtained during Cycles 21 -- 24, indicate that the upcoming Solar Maximum of Cycle 25 (SC25) is expected to be weaker than the current Cycle 24 (which nearing its end). The model results show that a deep extended solar activity minimum is expected during 2019 -- 2021, and that the next solar maximum will occur in 2024 -- 2025. The sunspot number at the maximum will be about 50 (for the v2.0 sunspot number series) with an error estimate of 15 -- 30~\%. The maximum will likely have a double peak or show extended periods (for 2 -- 2.5 years) of high activity. According to the hemispheric prediction results, SC25 will start in 2020 in the Southern hemisphere, and will have a maximum in 2024 with a sunspot number of about 28. In the Northern hemisphere the cycle will be delayed for about 1 year (with an error of $\pm 0.5$~year), and reach a maximum in 2025 with a sunspot number of about 23.
\end{abstract}

\section{Introduction}\label{introduction}
Growing interest in solar magnetic activity, and, in particular, in its prediction, is driven by the upcoming Solar Cycle 25 (SC25). As we approach the solar minimum of the current cycle, more predictions of the upcoming activity cycle are released. Predictions of the previous Solar Cycle 24 (SC24) were obtained with a wide range of methods and resulted in a wide range of predicted activity strengths, e.g., the predicted maximum sunspot number ranged from 42 to 197 \citep{Pesnell2012a}. Such diversity of the predicted cycle properties reflects limitations in our understanding of global processes in the Sun. 

The most successful empirical predictions of Solar Cycle 24 \citep[e.g,][]{Schatten2005} were made by assuming that the polar magnetic field strength during the preceding solar minimum represents the poloidal magnetic field which is later converted into the toroidal field by differential rotation. However, a dynamo model which is qualitatively capable of reproducing this phenomenon and predict solar-cycle evolution of the global magnetic field has not been developed.

The limitations of available models and poor knowledge of the current and past states of global magnetic field can be partially resolved by driving a model solution with observational data. For instance, \cite{Choudhuri2007} used the correlation between the strength of the polar fields during a solar minimum and the amplitude of the following activity cycle \citep{Svalgaard2005} to drive a 2D flux-transport model using observations of dipole moment parameters that indirectly characterize the polar field strength. The resulting forecast of SC24 was correct in the cycle strength but gave an incorrect time of the cycle maximum. Nevertheless, this indicates the potential of the data-driven methodology to improve the model predictions by including information from observations. Recently, \cite{Lemerle2017} suggested driving a model solution by an `emergence function', which describes emergence of magnetic flux on the solar surface. They predict that SC25 will be about about 20\% weaker than SC24 and that the mean sunspot number during the cycle maximum will be about 89 \citep{Labonville2019}. Surface flux-transport models have been used to advect magnetic fields obtained from synoptic magnetograms to predict the polar field strength and then use the empirical relation to estimate the next solar maximum \citep{Upton2014b}. According to this approach, the upcoming SC25 maximum will be similar to SC24 with an uncertainty of 15\% \citep{Upton2018}.

Recently, machine-learning techniques, in particular neutral network methods, have been employed to predict solar activity. For instance, statistics-based analysis of the sunspot area's butterfly diagram, trained on historical data sets, showed reasonable ability to reconstruct the evolution of the sunspot area for one solar cycle \citep{Covas2017}. However, the reconstructed longitudinal distribution is noisy, and the value of the sunspot area is underestimated. This approach suggests that SC25 will have a maximum sunspot number of 57$\pm$17 \citep{Covas2019}. Application of a neutral network algorithm to the flux-transport model and polar field observations led to the prediction that the upcoming solar cycle will be about 14\% weaker than the current cycle \citep{Macario-Rojas2018}.

Application of a Monte-Carlo technique to incorporate SOLIS/NSO and HMI/SDO synoptic magnetograms in a flux-transport model assuming the Babcock-Leighton dynamo mechanism \citep{Babcock1961,Leighton1969} in a flux-transport model demonstrated good agreement of the predicted poloidal field for up to one year \citep{Jiang2018}. In this research, the flux-transport model was driven by available magnetograms to obtain 50 realization of the future states. They predicted that SC25 will be about 10\% stronger (sunspot number $\sim 125$) than SC24 with probability of 95\% \citep{Jiang2018a}.
 
Because of model inaccuracies, as well as shortage and inaccuracies of observational data, realistic description of global solar dynamics is challenging. Modern data assimilation (DA) techniques allow us to link models with observations, taking into account uncertainties of both the models and data. 
The data assimilation methods are different from the data-driven approaches because they allow us to perform statistical cross-analysis of discrepancies between models  and observations and thereby obtain a more accurate description of current and past solar activity states (in our case, toroidal and poloidal fields and magnetic helicity) and estimate their uncertainties. This provides more realistic and consistent initial conditions for the prediction step. 

The Ensemble Kalman Filter (EnKF) method \citep{Evensen1994} has been used to predict SC24 by assimilating the sunspot number data into a non-linear mean-field dynamo model formulated in the form of a low-mode dynamical system \citep{Kitiashvili2008a,Kitiashvili2011}. It provided an accurate prediction of SC24 before its start in 2008. This methodology has been used to investigate the possibility of performing forecasting at various stages of the solar activity cycles \citep{Kitiashvili2016}. It was found that prediction of a future solar cycle is possible only using initial model states corresponding to either the preceding solar minimum or the period of polar field reversals. During these time periods the global magnetic field structure is dominated either by poloidal or toroidal field components. This reduces uncertainties in the initial model state for the prediction solution.
Nevertheless, the accuracy of those predictions initiated during polar field reversals decreases with increased time-lag between the polar field reversals in the two hemispheres. Another variational data assimilation approach, 4DVar, is more challenging than the sequential EnKF method for application to actual observations because of its complexity but can provide estimates of model parameters, such as kinetic helicity \citep{Jouve2011}.
 
Many existing mean-field and flux-transport dynamo models \citep[e.g.][]{Dikpati2007, Wang1991,Cameron2007,Choudhuri2007,Kitiashvili2009,Pipin2015a,Jiang2013a,Karak2013,Upton2014b,Wang2016} are capable of capturing various manifestations of the global evolution of the magnetic fields. However, such complex two- and three-dimensional models require more complete sets of observations for the assimilation procedure than are currently available in long-term records of solar activity, as well as proper calibration of numerous model properties.

In our previous work, the Ensemble Kalman Filter (EnKF) method has been applied for assimilation of the sunspot number series into the low-order Parker-Kleeorin-Ruzmaikin (PKR) mean-field dynamo model for reconstruction of past solar activity and prediction of SC24 \citep{Kitiashvili2008a,Kitiashvili2009,Kitiashvili2011,Kitiashvili2016,Kitiashvili2019}. The annual sunspot number series was converted into the mean toroidal magnetic field following the suggestion of \cite{Bracewell1988}. The next step is to include information about the global surface evolution of magnetic fields, which is currently available from 42~years of synoptic observations.
In this paper, synoptic magnetograms are used to study the capability of the data assimilation method to predict annual variations of the magnetic field components and the sunspot number, as well as the evolution of solar activity separately in each hemisphere. 

We start from a brief description of the Parker-Kleeorin-Ruzmaikin dynamo model (Section~\ref{PKR}), observational data, and a procedure for decomposition of the synoptic magnetograms into magnetic field components (Section~\ref{Observations}). The Ensemble Kalman Filter method is described in Section~\ref{EnKF}. Section~\ref{SC} presents test results for reconstruction of solar activity during SC23 and 24. Our forecast for SC25 based on assimilation of synoptic magnetograms is presented in Section~\ref{SC25}. The paper concludes with a discussion of results and the influence of data limitations and uncertainties on predictive capability and accuracy (Section~\ref{discussion}).

\section{Parker-Kleeorin-Ruzmaikin dynamo model}\label{PKR}
We employ the mean-field dynamo model previously formulated by \cite{Kitiashvili2008a,Kitiashvili2009}. It combines Parker's migratory  $\alpha$-dynamo model \citep{Parker1955} with the equation of the magnetic helicity balance \citep{Kleeorin1982,Kleeorin1995} and describes the coupling between turbulence and magnetic fields:  
\begin{eqnarray}
\frac{\partial A}{\partial t}&=&\alpha B+\eta \nabla ^{2}A \nonumber \\
\frac{\partial B}{\partial t}&=&G\frac{\partial A}{\partial x}+\eta
\nabla ^{2}B,\\
\frac{\partial \alpha_{m}}{\partial t}&=&-\frac{\alpha_{m}}{T}+\frac{Q}{4\pi \rho}\left[\left<{\bf B}\right>  \cdot
(\nabla \times \left<{\bf B}\right>)-\frac{\alpha}{\eta}\left<{\bf B}\right>^{2}\right], \nonumber
\end{eqnarray}
where $A$ and $B$ are the vector-potential and the toroidal field component of magnetic field $\rm{\bf B}$, $\alpha=\alpha_{h}+\alpha_{m}$ is total helicity, which includes the hydrodynamic ($\alpha_{h}$), and magnetic  ($\alpha_{m}$) parts, $\eta=\eta_{t}+\eta_{m}$ (where usually $\eta_{m} \ll \eta_{t}$) is total magnetic diffusion, $\eta_{t}$ is the turbulent and $\eta_{m}$ is the molecular magnetic diffusivity, $G$ describes the rotational shear, $\rho$ is density, $Q$ and $T$ are coefficients which describe characteristic scales and times, and $t$ is time.

Following the suggestion of \cite{Weiss1984}, the dynamo model equations (Eq. 1) are converted into a non-dimensional dynamical system by averaging the model properties in the radial direction. Taking into account a single Fourier mode propagating in the equatorward direction \citep{Kitiashvili2009} the initial set of equations has the following form
\begin{eqnarray}
\frac{{\rm d} A}{{\rm d} t}&=& D B- A, \nonumber \\
\frac{{\rm d} B}{{\rm d} t}&=&{\rm i} A - B,  \\
\frac{{\rm d} \alpha_{m}}{{\rm d} t}&=&-\nu \alpha_{m} - D \left[
 B^{2}-\lambda A^2 \right], \nonumber
\end{eqnarray}
where $A$, $B$, $\alpha_{m}$, and $t$ are non-dimensional variables, $D=D_{0}\alpha$ is a non-dimensional dynamo number, $D_{0}=\alpha_{0}Gr^{3}/\eta^{2}$, $\alpha_{0}=2Qk\upsilon_{A}^{2}/G$, $\upsilon_{A}$ is the Alfv\'en speed, $\lambda=(k^2\eta/G)^2=Rm^{-2}$, $k$ is a characteristic wavelength, and $\nu$ is the ratio of characteristic turbulence time-scales. The sign of the dynamo number gives the direction of the dynamo waves migration in latitude: from higher to lower latitude if the dynamo number is negative, and the opposite for positive $D$. We consider only negative dynamo number to make the model consistent with the butterfly diagram. The reduced dynamo model describes the evolution of three basic properties: the mean global toroidal and poloidal field components and magnetic helicity. 

The dynamical system (Eg. 2) can be decomposed into real and imaginary parts as $A=a_1+ia_2$, $B=b_1+ib_2$, $\alpha_m=c_1+ic_2$, where the dynamo number $D=\alpha D_0$, total helicity is $\alpha=\alpha_h+\alpha_m$ with quenching $\alpha_h=\alpha_k/(1+q (b_1^2+b_2^2))$. Then, the dynamical system can be rewritten in the following form
\begin{eqnarray} 
\dot{a_1}&=& - a_1 + D_0 (\alpha_h b_1 + b_1 c_1 - b_2 c_2) \nonumber\\
\dot{a_2}&=& - a_2 + D_0 (\alpha_h b_2 + b_1 c_2 + b_2 c_1) \nonumber\\
\dot{b_1}&=& - a_2 - b_1 \\
\dot{b_2}&=& a_1 - b_2 \nonumber\\ 
\dot{c_1}&=& - \nu c_1 + D_0 (\alpha_h + c_1) [\lambda (a_1^2-a_2^2) - (b_1^2-b_2^2)] - 2 D_0 c_2 (\lambda a_1 a_2  - b_1 b_2) \nonumber\\
\dot{c_2}&=& - \nu c_2 + 2 D_0 (\alpha_h + c_1) (\lambda a_1 a_2 - b_1 b_2) + D_0 c_2 [\lambda (a_1^2-a_2^2) - (b_1^2-b_2^2)].\nonumber
\end{eqnarray}

As was demonstrated before \citep[e.g.][]{Kitiashvili2008a,Kitiashvili2009}, the PKR model periodic solution qualitatively reproduces the basic properties of solar activity cycles, such as the Waldmeier rules, the mean shape of the cycles, etc. In this work, we apply this model to available synoptic magnetograms and investigate the model's predictive capabilities for both the magnetic field components and the sunspot number in each hemisphere of the Sun. 
The model parameters:  $D_0=-3.3$, $\alpha_k=1.8$, $\nu=0.013$, $q=5.87$, and $\lambda=3.2 \times 10^{-6}$, as well as the initial conditions, were chosen by performing trial runs to obtain a periodic solution, which qualitatively reproduces the observed cyclic evolution of the global magnetic field. 

\section{Observations: Synoptic magnetograms}\label{Observations}
\begin{figure}[h]
	\begin{center}
		\includegraphics[scale=1]{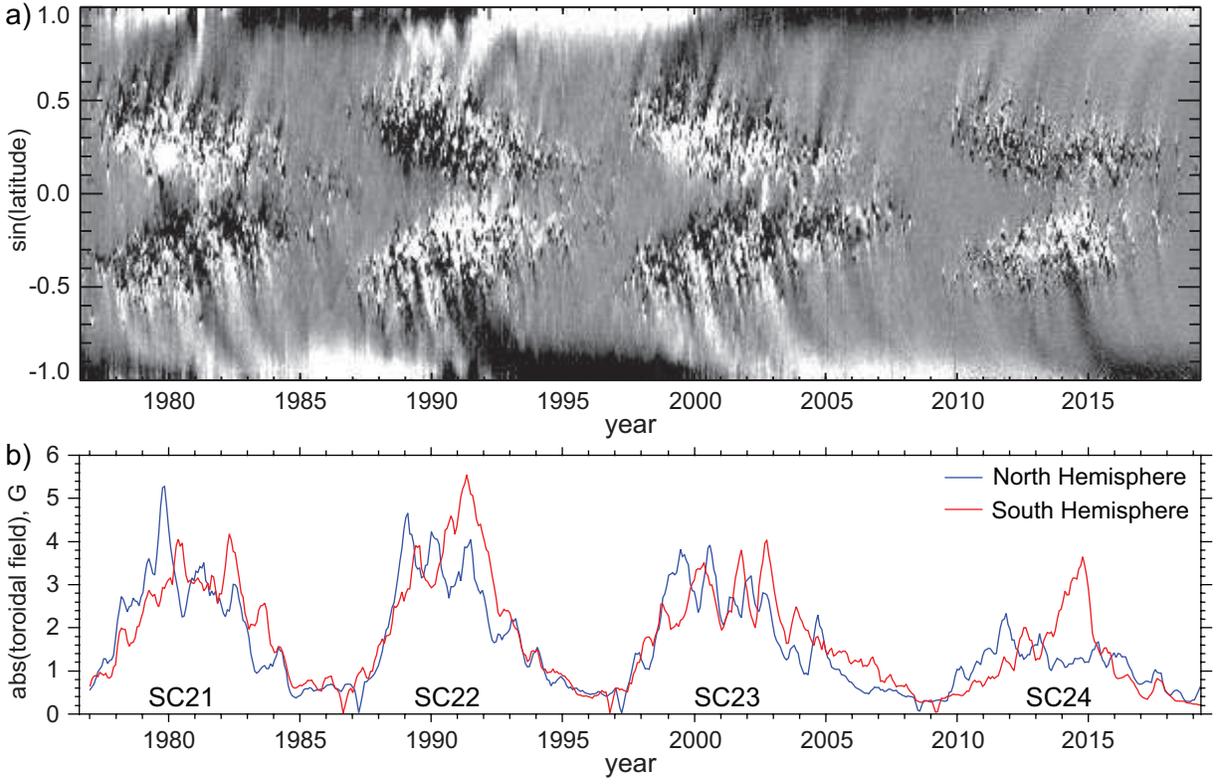}
	\end{center}
	\caption{Panel a: Synoptic magnetogram covering four solar cycles from 1976 to 2019. The grey color-scale is saturated at range $\pm10$~G. Panel b: Temporal variations of the mean unsigned toroidal magnetic field component in the Northern (blue curve) and Southern (red) hemispheres. \label{fig:SynMagnetogramm}}
\end{figure}

In the current work, we use synoptic magnetograms for the last four solar cycles (Fig.~\ref{fig:SynMagnetogramm}a), which were obtained for the period from 1976 (Carrington rotation 1645) to 2019 (Carrington rotation 2216) from Kitt Peak Observatory \citep{Harvey1980,Worden2000}, the SOLIS instrument \citep{Keller2003}, and SOHO/MDI and SDO/HMI \citep{Scherrer1995,Scherrer2012}. The synoptic magnetograms have been reduced to the original KPO spatial resolution of $360\times180$~pixels and to only the radial magnetic field component. Magnetic field measurements unavailable in the polar regions were copied from lower latitudes. Synoptic magnetograms for Carrington rotations 1854, 2015, 2016, 2040, and 2041 are not available, and to avoid gaps in the data, the closest magnetograms have been used.

Decomposition of the synoptic magnetograms into toroidal and poloidal field components is a challenging task due to the difficulty of finding a unique solution. To simplify the magnetic field decomposition problem we assume that the high-latitude magnetic field (above the active latitudes) characterizes the poloidal field component and that the unsigned flux in the active latitudes corresponds to the toroidal field. This assumption is acceptable for the 1D model with some level of uncertainty, because it requires estimates of the relative behavior of the field components. 

\begin{figure}[h]
	\begin{center}
		\includegraphics[scale=1]{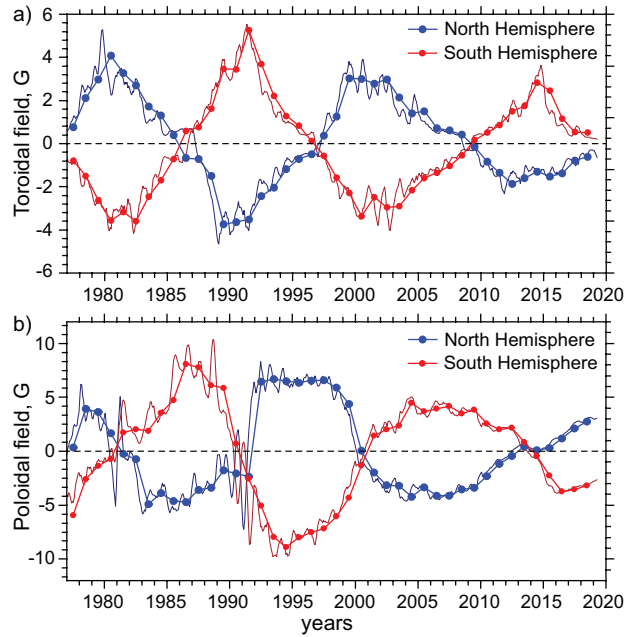}
	\end{center}
	\caption{Temporal variations of the toroidal (panel a) and poloidal fields (panel b) in the Northern (blue curves) and Southern hemispheres (red curves). Circles indicate magnetic field values used in the analysis. \label{fig:TorPolSN}}
\end{figure}

To account for toroidal field reversals, the sign of the estimated toroidal field is prescribed according to the Hale polarity law. Figure~\ref{fig:SynMagnetogramm}b shows variations of magnitude of the estimated toroidal field with time for each hemisphere. The time-series of the estimated toroidal and poloidal fields are averaged over 1-year intervals and are shown by circles for each hemisphere in Figure~\ref{fig:TorPolSN}. Thin curves show unsmoothed variations of the fields for reference. 

\begin{figure}[h]
	\begin{center}
		\includegraphics[scale=0.9]{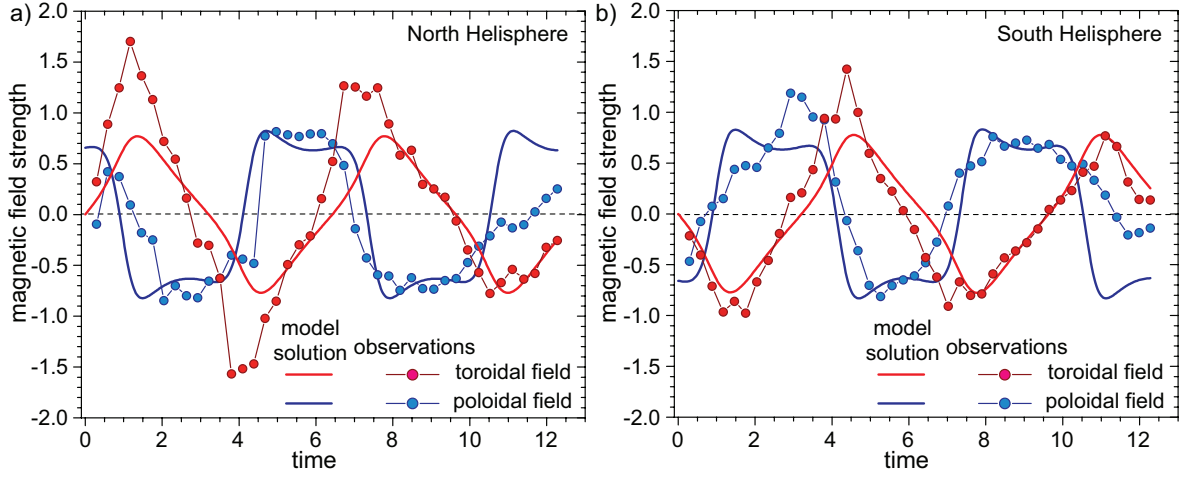}
	\end{center}
	\caption{Time-series of the annual toroidal (red dots) and poloidal (blue) field observations calibrated to the corresponding periodic dynamo solutions (thick curves) for the Northern (panel a) and Southern hemispheres (b). The magnetic fields and time units are non-dimensional. \label{fig:MFCalibr}}
\end{figure}

The resulting annual observations have been normalized to match the model periodic solutions for the toroidal and poloidal fields (Fig.~\ref{fig:MFCalibr}) for each hemisphere. Normalization for the poloidal field was chosen for a best agreement for the field amplitude. For the toroidal field, the normalization is performed relative to the last observed solar cycle, following the approach of \cite{Kitiashvili2008a}. Figure~\ref{fig:MFCalibr}b shows an example of the toroidal component of the magnetic field calibration in the model solutions for the prediction of SC25. Traditionally, solar activity cycles are characterized by the sunspot number; the toroidal field can be converted to the sunspot number with a corresponding normalization. Comparison of the observed hemispheric sunspot number and that estimated from the synoptic magnetograms is shown in Figure~\ref{fig:MagSN}.

\section{Data assimilation methodology: Ensemble Kalman Filter}\label{EnKF}
The problem of forecasting the behavior of a physical system with multiple interacting and evolving non-linear processes is very common in many fields, such as climate change, atmosphere and ocean dynamics, and others. Discrepancies between model solutions and observations significantly restrict or even prevent building reliable forecasts. The origin of these discrepancies is incompleteness of the models, as well as shortage and uncertainties of observations. In such situations, observations and models can be linked through a cross-analysis of measurements and model solutions together with estimation of errors and uncertainties. This mathematical procedure, called {\it Data Assimilation}, represents a wide class of methods \citep[see e.g.,][]{Evensen1997,Kalnay2002}. 

In this paper, we use the Ensamble Kalman Filter method \citep[EnKF,][]{Evensen1997}, which is an extension of the Kalman Filter method \citep{Kalman1960} for non-linear systems. Because of errors and uncertainties, observations are described in terms of the true state of solar global activity, $\psi^t$, which in our case is represented by the evolution of the magnetic field components and magnetic helicity. Measurements, $m$, can be described by the relation $m=M\psi^t+\epsilon$, where $M$ is the measurement functional, describes the relationship between the model properties and observational errors, $\epsilon$. Since the true state of global solar activity is unknown, observational data is considered as an ensemble of possible measurements, $m_j$: 
\begin{eqnarray} 
m_j=M\psi^t+\epsilon_j, 
\end{eqnarray} 
where $j=1, ..., N$, and $N$ is the number of ensemble members. The ensemble error covariance matrix of observations is then $C^e_{\epsilon\epsilon}=\overline{\epsilon\epsilon^T}$. In this work, all analysis is performed for $N=300$.

\begin{figure}[h]
	\begin{center}
		\includegraphics[scale=1]{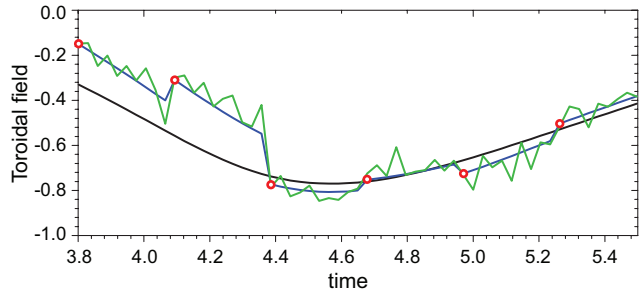}
	\end{center}
	\caption{Illustration of the EnKF analysis procedure of a model solution correction (thick black curve) according to observations (circles) to build a forecast (green dashed curve).  Blue curve shows an initial guess of the field variations, and green curve shows a best estimate of the field, obtained by the EnKF analysis. \label{fig:assim}}
\end{figure}

The physical model can be presented in the form \citep{Evensen1997}:
\begin{eqnarray} 
{\rm d \psi}=G(\psi)dt+h(\psi)dq,
\end{eqnarray} 
where $G(\psi)$ is a model operator, the term $h(\psi)dq$ describes the model errors, in which $q$ represents random variations, $h(\psi)$ is the uncertainty of the model state, and $t$ is time. 
In this context, the model solution can be considered as a predicted state ($\psi^f$) of the global solar activity for each moment of time (thick black curve, Fig.~\ref{fig:assim}), described as a combination of the `true' state  and errors, $\psi^f=\psi^t+e^f$, with an error covariance $C^f_{\psi\psi}=\overline{(\psi^f - \psi^t)(\psi^f - \psi^t)^T}$. Because the `true' state is not known, the error covariance for the ensemble of states becomes $(C^e_{\psi\psi})^f=\overline{(\psi^f - \overline{\psi^f})(\psi^f - \overline{\psi^f})^T}$, and allows us to compute the Kalman gain as 
\begin{eqnarray}
K^e=\frac{(C^e_{\psi\psi})^f M^T}{M (C^e_{\psi\psi})^f M^T + C^e_{\epsilon\epsilon}}
\end{eqnarray} 
to obtain a first guess for the corrected model solution (blue curve, Fig.~\ref{fig:assim}). Such estimates are obtained sequentially for each observation, which is considered as an ensemble of possible states for the unknown true state. This analysis has been used to compute a best estimate for the likelihood solution (green curve in Fig.~\ref{fig:assim}) in the following form
\begin{eqnarray}
\psi^a_j=\psi^f_j+K^e (d_j-M\psi^f_j).
\end{eqnarray} 

The practical implementation of the EnKF method to the solar activity prediction problem has been performed in three steps, previously described in detail by \cite{Kitiashvili2011}: 1) preparation of the observational data, 2) assimilation for past activity states, and 3) forecast. 

Step 1 includes analysis of the synoptic magnetograms collected from different space and ground instruments, by performing decomposition of the magnetograms into toroidal and poloidal field components, and recalibration of the observational data as described in Section~\ref{Observations}. To separate analysis for the past and future activity, two phases are identified: `analysis' (corresponding to Step 2) and `prediction' (Step 3). Each of these phases is indicated in all figures showing the results of this study. 

\section{Test predictions of past and current solar activity}\label{SC}
\subsection{Application of the toroidal and poloidal fields to solar cycle forecasts}\label{TorPol}
In this section we describe testing of the predictive capabilities of assimilation of toroidal and poloidal field observations into the PKR dynamo model using the EnKF method. The goal is to reproduce the evolution of the magnetic field components for each hemisphere, as well as the sunspot number, during SC23 and SC24 based on the data for previous cycles.

\subsubsection{Solar Cycle 23 test prediction}\label{SC23}
\begin{figure}[h]
	\begin{center}
		\includegraphics[scale=0.9]{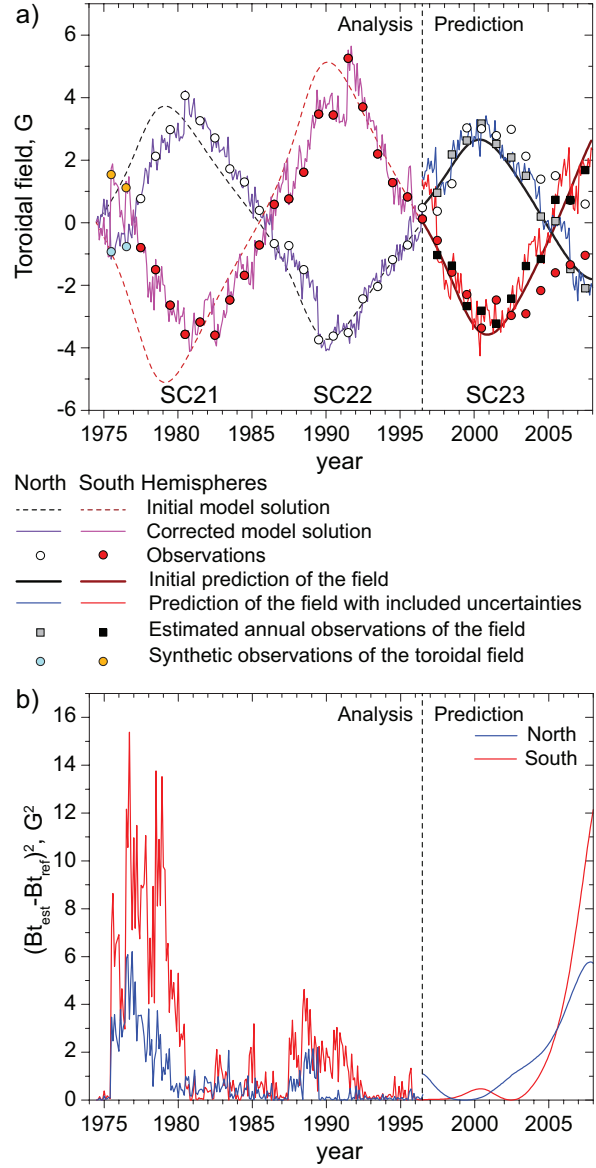}
	\end{center}
	\caption{Panel a: Evolution of the mean toroidal field in the Northern and Southern hemispheres based on field observations for SC21 and SC22, prediction of the mean toroidal field component variation during SC23, and comparison of the prediction with the toroidal field observations. Panel b shows errors of the initial periodic solutions relative to the actual observations during data assimilation analysis and prediction steps. Blue lines show errors for the Northern and red lines for Southern hemispheres. Vertical dashed lines indicate the prediction start time. \label{fig:SC23Predc2TorNSBerr}} 
\end{figure}

\begin{figure}[h]
	\begin{center}
		\includegraphics[scale=0.7]{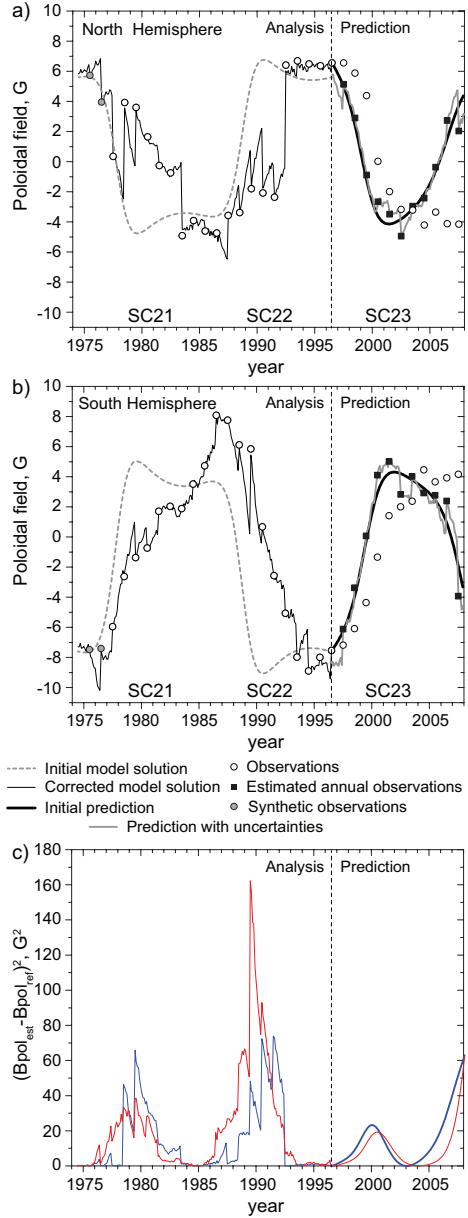}
	\end{center}
	\caption{Evolution of the mean poloidal field in the Northern (panel a) and Southern (panel b) hemispheres based on field observations of SC21 and SC22, prediction of the mean poloidal field variation during SC23, and comparison of the prediction with the poloidal field observations. Panel c shows errors of the initial periodic solutions (dashed curves) relative to the actual observations (empty circles) during data assimilation analysis and prediction steps. Blue lines show errors for the Northern and red lines for Southern hemispheres. Vertical dashed lines indicate the prediction start time. Blue lines correspond to errors for Northern and red for Southern hemispheres. Vertical dashed lines indicate prediction start time. \label{fig:SC23Predc2PolNSSAerr}}
\end{figure}

Reconstruction of global solar activity during SC23 has been tested using synoptic magnetograms obtained during two previous cycles, SC21 and SC22 (section~\ref{Observations}) from 1977.5 to 1996.5. Because the available observations start from the rising phase of SC21, to assimilate the global field evolution corresponding to the solar minimum between SC20 and SC21, two synthetic observations for 1976.5 and 1975.5 have been added to the observational time-series. Figure~\ref{fig:SC23Predc2TorNSBerr}a shows the annual observational data (circles) and the model solution for the toroidal field (dashed curves) normalized to the SC22 maximum data for each hemisphere. Also, the model solution phase is chosen to match the phase of the toroidal field of SC22. The start of the prediction phase is indicated by the vertical dashed line (Fig.\ref{fig:SC23Predc2TorNSBerr}a). Using the EnKF procedure, the periodic model solution (dashed curves) is corrected according to the annual observations for the toroidal (thin red and blue curves, Fig.\ref{fig:SC23Predc2TorNSBerr}a) and poloidal fields (thin black curves, Fig.\ref{fig:SC23Predc2PolNSSAerr}a,b) for the corresponding hemispheres. The additional model variables, e.g. magnetic helicity, for which observations are not available, are generated from the model solution with imposed noise of 10\%. 

Comparison of the model prediction with the actual toroidal field variations shows good agreement for both hemispheres up to the SC23 maximum (Fig.~\ref{fig:SC23Predc2TorNSBerr}b). After the maximum, the predicted toroidal field in the Northern hemisphere quickly deviates from the observed evolution. In the Southern hemisphere, deviations of the predicted toroidal field become significant two years after the SC23 maximum. 
Figure~\ref{fig:SC23Predc2PolNSSAerr}a,b shows results of analysis for the mean poloidal field for each hemisphere. The predicted evolution of the poloidal field in both hemispheres quickly deviates from the actual data after the first year after prediction start. The predicted time of the field reversals for both hemispheres is one year earlier than the actual one. Nevertheless, the maximum strength of the poloidal fields is estimated correctly for both hemispheres. The predicted time of the strongest poloidal field strength is two years earlier for the Northern hemisphere and about four years earlier for the Southern hemisphere. 
The origin of the prediction discrepancies is likely due to a significant phase deviations between the model solution (dashed curves in Fig.~\ref{fig:SC23Predc2PolNSSAerr}a,b) and the observations (empty circles).

\begin{figure}[h]
	\begin{center}
		\includegraphics[scale=0.9]{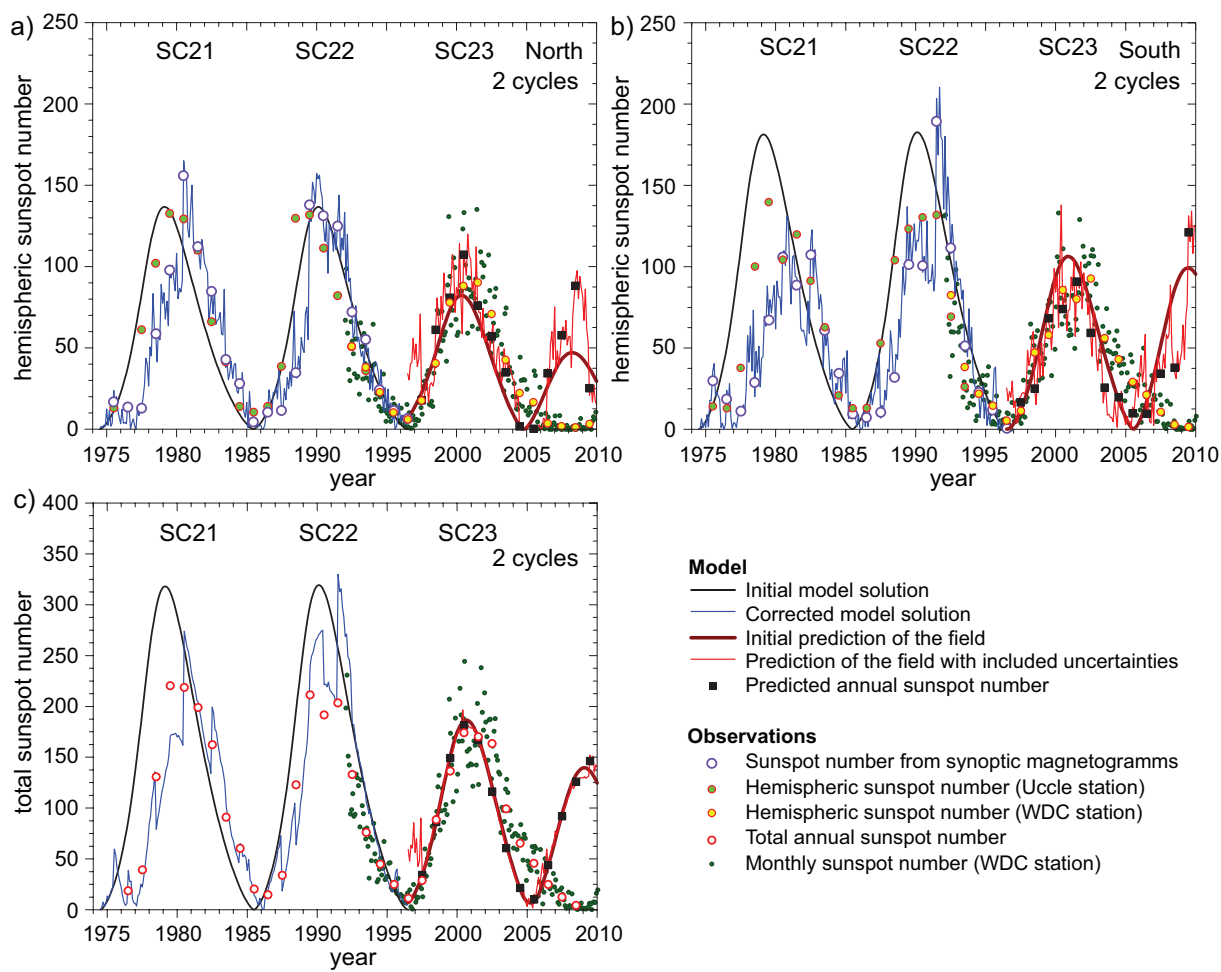}
	\end{center}
	\caption{Reconstruction of sunspot number variations during SC23 for the Northern hemisphere (panel a), Southern hemisphere (panel b), and the total sunspot number (panel c). Solid black curves show an initial periodic model solution, blue curves correspond to the corrected solution according to the observed fields, thick brown curves represent the initial forecast of the sunspot number variations, and red thin curves show the prediction estimates taking into account data uncertainties and model errors. Black squares show estimates for the annual sunspot number, other symbols show the actual observational data. Green dots show the monthly sunspot number and are given for reference. \label{fig:SC23c2SN}}
\end{figure}

Because the toroidal magnetic field correlates with the sunspot number, the results can be presented in terms of sunspot number variations (Fig.~\ref{fig:SC23c2SN}). The sunspot number prediction for SC23 is in good agreement with the actual hemispheric sunspot number data for the whole cycle in the Northern hemisphere (Fig.~\ref{fig:SC23c2SN}a). For the Southern hemisphere, the prediction results are in agreement with the amplitude and time of the solar activity maximum. Deviations between the predicted and actual observations gradually increase in the declining phase of the cycle (Fig.~\ref{fig:SC23c2SN}b). The total sunspot number (panel c) is correctly reconstructed for most of the cycle.

\subsubsection{Solar Cycle 24 test prediction}\label{SC24}
Test prediction for Solar Cycle 24 is performed by using assimilation of  observational data for the previous two and three solar cycles. Utilization of only two cycles can be considered as an additional test to investigate the limitations of very short observational time-series. 

\begin{figure}[h]
	\begin{center}
		\includegraphics[scale=0.9]{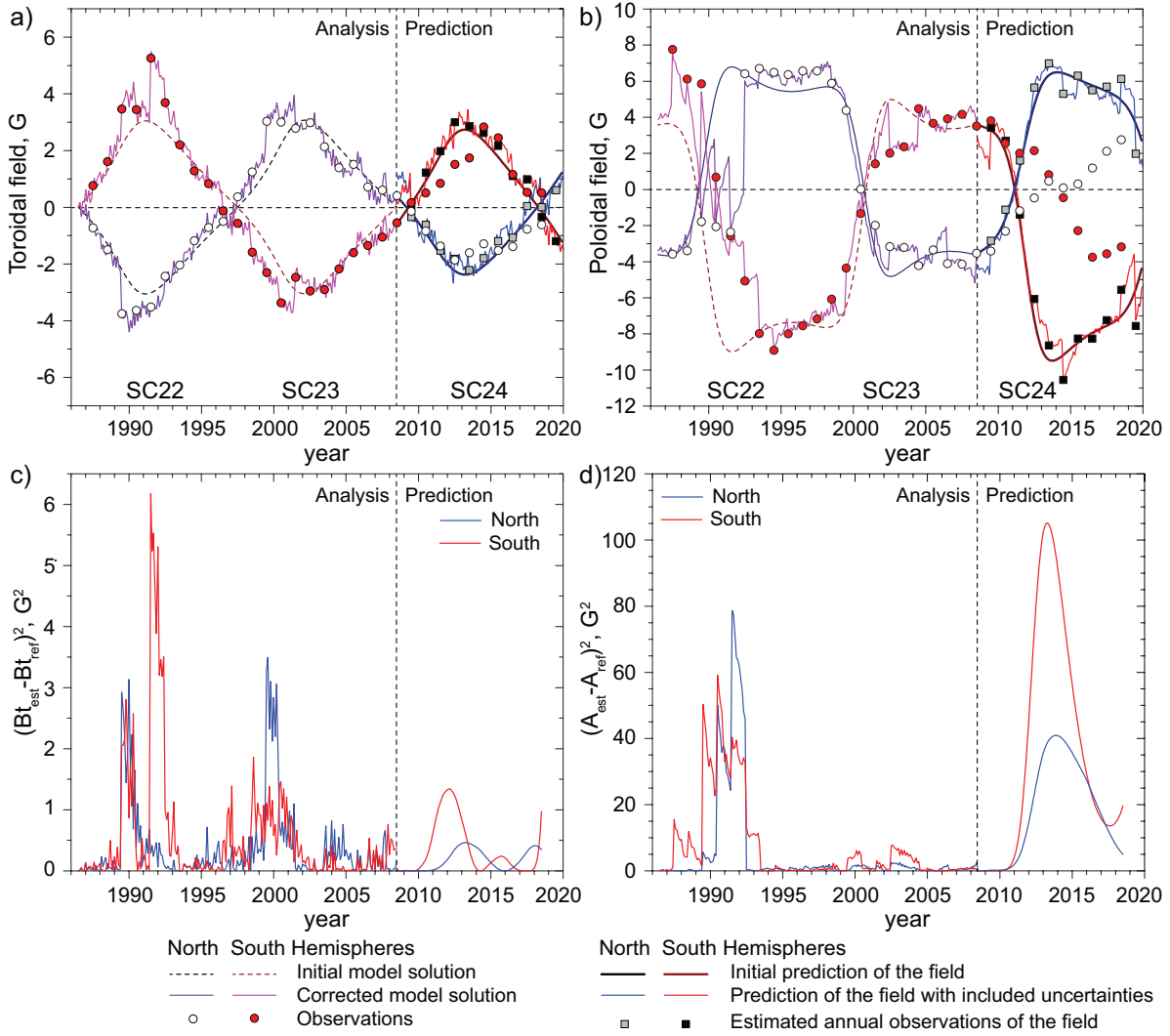}
	\end{center}
	\caption{Evolution of the mean toroidal (panel a) and poloidal (b) fields in the Northern and Southern hemispheres based on field observations for SC22 and SC23 (case 1), and prediction of the mean toroidal and poloidal field components during SC24. Panels c) and d) show deviations of the model solutions for the magnetic field components from the actual observational data. Blue curves correspond to the errors for the Northern hemisphere, and the red curves for the Southern hemisphere. Vertical dashed lines indicate the prediction start time.\label{fig:SC24c2SN_TorPol}}
\end{figure}

\subsubsubsection{Case 1: Assimilation of data for two solar cycles}
In this case, we use the available synoptic magnetograms from 1987.5 to 2008.5 for prediction of SC24. The phase of the periodic dynamo solution fits the activity phase very well, which allows us to make a good prediction for the toroidal field for the whole solar cycle in the Northern hemisphere (Fig.~\ref{fig:SC24c2SN_TorPol}a,c). In the Southern hemisphere, despite growing discrepancies between the estimated and actual toroidal field variations, the cycle duration is predicted correctly. The poloidal field prediction has good agreement with the observations for the first two years in both hemispheres, and then it significantly deviates from the observations (Fig.~\ref{fig:SC24c2SN_TorPol}b,d). The forecast of the sunspot number (Fig.~\ref{fig:SC24c2SN}) is in agreement with observations. Most of the discrepancies are related to the complicated shape of the cycle near its maximum in the Northern hemisphere (panel a). Such relatively short-time variations are not described in the current model formulation. The estimates of the total sunspot number correctly predict the rise and decay rate, but the strength of the activity cycle is underestimated by 20\% (Fig.~\ref{fig:SC24c2SN}c).  

\begin{figure}[h]
	\begin{center}
		\includegraphics[scale=0.9]{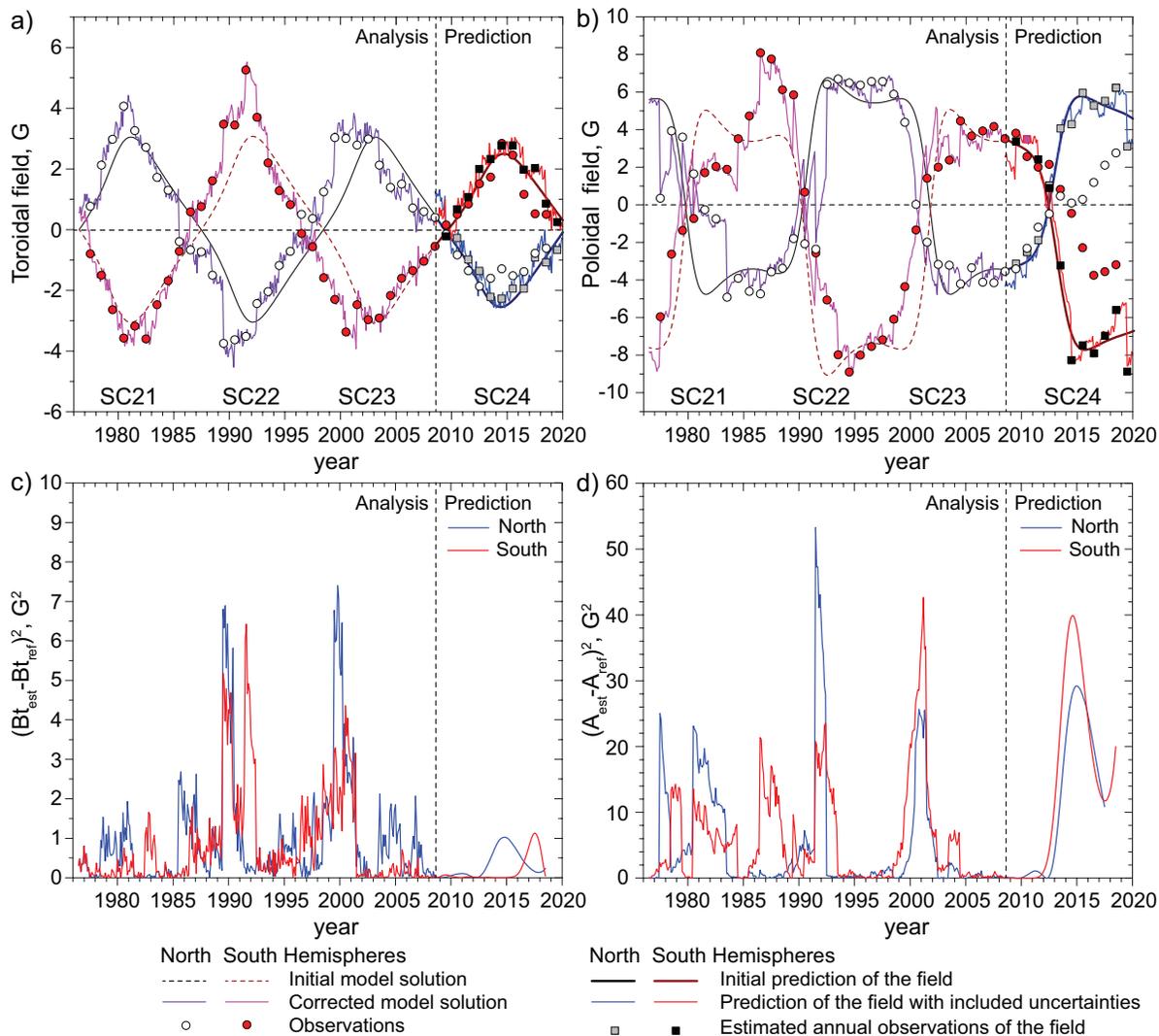}
	\end{center}
	\caption{Evolution of the mean toroidal (panel a) and poloidal (b) fields in the Northern and Southern hemispheres based on the field observations for three solar cycles (case 2), and prediction of the mean toroidal and poloidal field components variation during SC24. Panels c) and d) show deviations of the model solutions for the magnetic field components from the actual observational data. Blue curves correspond to the errors for the Northern hemisphere, and the red curves for the Southern hemisphere. Vertical dashed lines indicate the prediction start time. \label{fig:SC24c3NS_TorPol}}
\end{figure}

\begin{figure}[h]
	\begin{center}
		\includegraphics[scale=0.9]{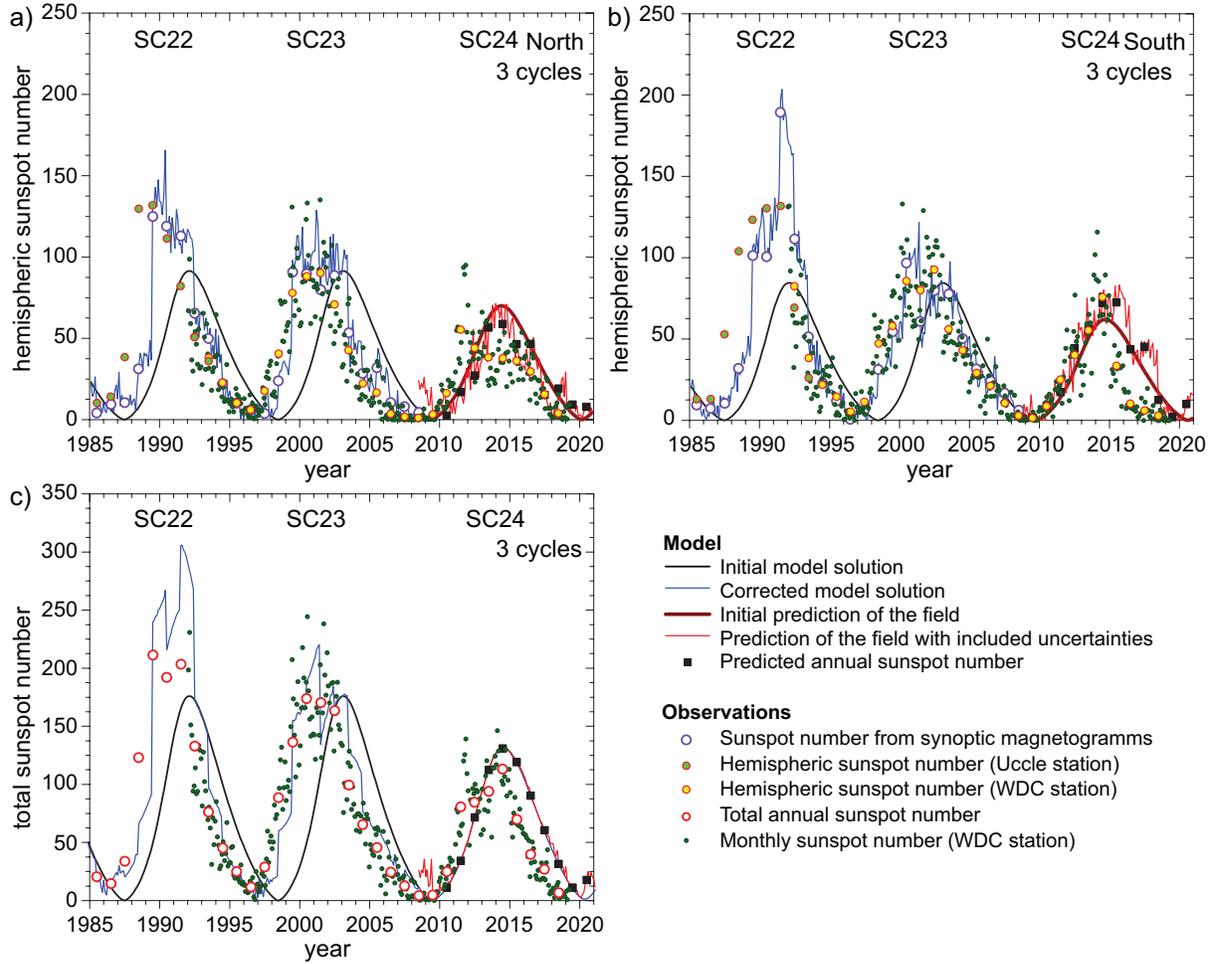}
	\end{center}
	\caption{Prediction of the sunspot number variations in SC24 for the Northern (panel a), Southern hemisphere (b), and the total sunspot number (panel c) assuming availability of observations for three previous cycles (case 2). \label{fig:SC24c3SN}}
\end{figure}

\subsubsubsection{Case 2: Assimilation for three solar cycles}
In this test case, we use the available magnetic field measurements from 1977.5 to 2008.5 for prediction of SC24. Results of the analysis are presented in Figures~\ref{fig:SC24c3NS_TorPol} and~\ref{fig:SC24c3SN}. The predicted evolution of the toroidal field is in good agreement for both hemispheres, although there are some discrepancies during the solar maximum in the Northern hemisphere. 
Increasing discrepancies between the predicted and observed toroidal fields during the decay phase of solar activity in the Southern hemisphere are expected because of the step-like variations of the predicted toroidal field evolution (red thin curve, Fig.~\ref{fig:SC24c3NS_TorPol}a) and the sunspot number (red thin curve, Fig.~\ref{fig:SC24c3SN}b). This behavior indicates accumulation of errors during the analysis and, in general, gives us a warning that the forecast quality is potentially low. This effect previously was discussed by \cite{Kitiashvili2008a} for assimilation of sunspot number time-series. 

Including the additional solar cycle in the assimilation procedure improves the forecast for the poloidal fields in both hemispheres (Fig.~\ref{fig:SC24c3NS_TorPol}b,d). Accuracy of the prediction is good for up to 3 years and provides a correct prediction for the time of the polar field reversals. After this, the predicted and observed field components quickly diverge. The sunspot number estimates (Fig.~\ref{fig:SC24c3SN}) show good agreement with the actual data for both hemispheres. Some deviations in the shape of the predicted activity cycles are expected, and this reflects restrictions of the dynamo model formulation. The total sunspot number maximum is slightly overestimated, but in general the prediction results show a good agreement for the whole solar cycle.

\subsection{Effects of poloidal field observations on the solar activity forecast}\label{onlyTor}
\begin{figure}[h]
	\begin{center}
		\includegraphics[scale=0.9]{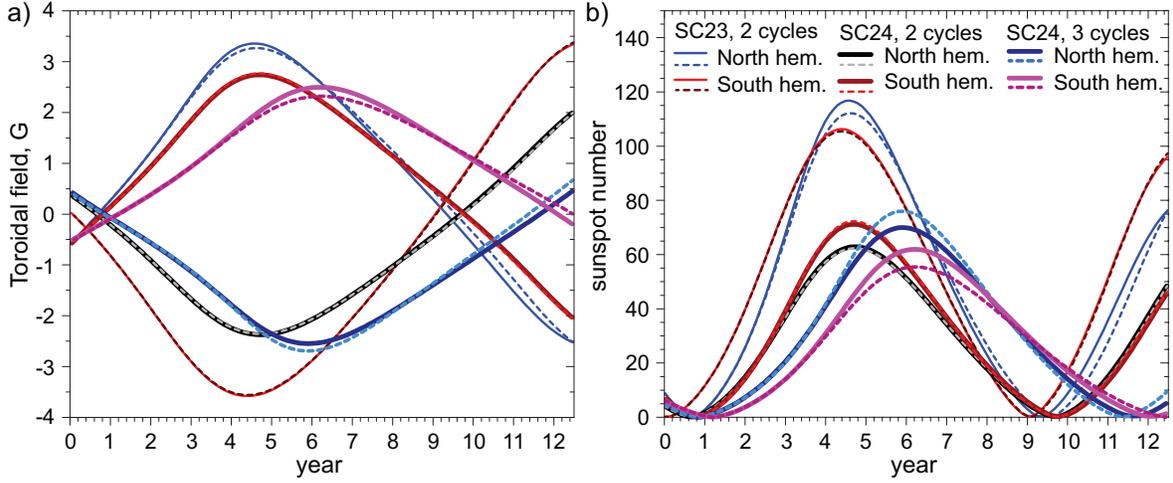}
	\end{center}
	\caption{Comparison of the first guess predictions for the toroidal field (panel a) and sunspot number (panel b) for two cases: (1) when only toroidal field observations are assimilated (dashed curves), and (2) when both the toroidal and poloidal fields are used (solid curves). The tests are performed for reconstruction of SC23 and SC24. Time $t=0$ correspondts to the prediction start time. \label{fig:BtW_comparison}}
\end{figure}

Our previous predictions of solar activity cycles \citep{Kitiashvili2008a,Kitiashvili2016} have been performed using sunspot number data and assuming that the sunspot number correlates with the toroidal field \citep{Bracewell1988}. In those studies, the poloidal field was generated from the model solution and perturbed with noise. The synoptic magnetograms available for the last four cycles of solar activity allow us to more accurately estimate global variations of the toroidal and poloidal fields and their uncertainties (section~\ref{Observations}). For testing how the poloidal field contribution improves the forecasts of SC23 and SC24, we replace the poloidal field observational data with synthetic observations generated by the dynamo model. Figure~\ref{fig:BtW_comparison} shows comparison of  the predicted toroidal field variations and the sunspot number for both hemispheres when only the toroidal field data were used (dashed curves) and for the case when both toroidal and poloidal field data were assimilated (solid curves). For both hemispheres the strongest deviations takes place at the solar maximum (Fig.~\ref{fig:BtW_comparison}). The difference in the sunspot cycle maximum predictions reaches 10.5\% for SC24 in the Southern hemisphere using the observations of 3 cycles and --8.6\% in the Northern hemisphere (Table~\ref{table}). Other tests showed significantly weaker discrepancies. Nevertheless, these test results indicate that the poloidal field measurements can be very important for improving prediction accuracy.  

\section{Solar Cycle 25 prediction}\label{SC25}
\begin{figure}[h]
	\begin{center}
		\includegraphics[scale=0.9]{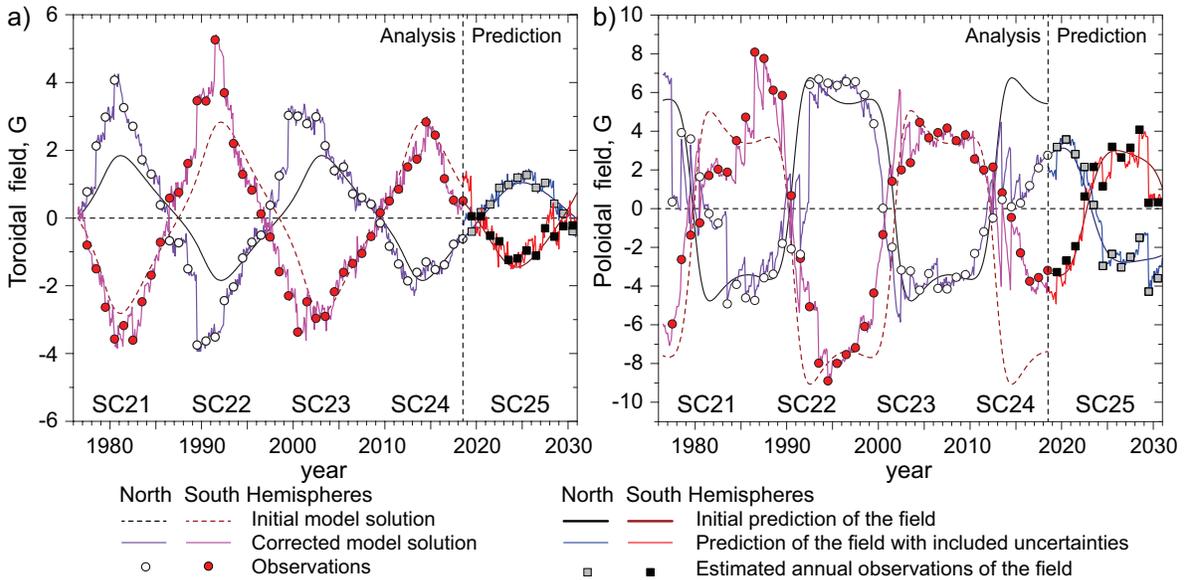}
	\end{center}
	\caption{Prediction for the  mean toroidal (panel a) and poloidal (b) fields in the Northern and Southern hemispheres based on field observations for three solar cycles. Vertical dashed lines indicate the prediction start time. \label{fig:SC25c4predTorPol}}
\end{figure}

\begin{figure}[h]
	\begin{center}
		\includegraphics[scale=0.9]{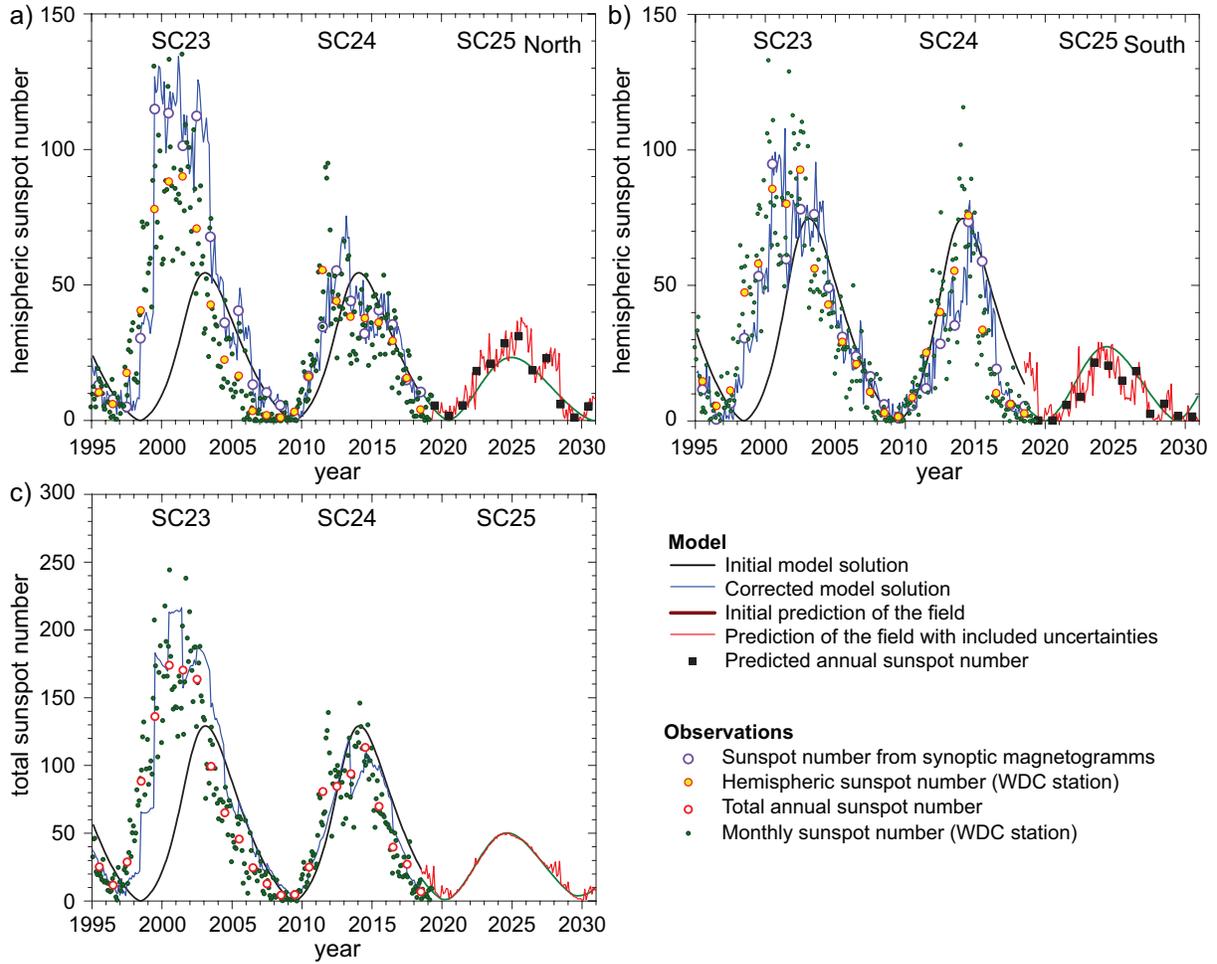}
	\end{center}
	\caption{Prediction of the sunspot number variations for SC25 for the Northern (panel a), Southern hemisphere (b), and the total sunspot number (panel c) using the synoptic magnetograms for the last four solar cycles. \label{fig:SC25Pred4cyclesTPNS}}
\end{figure}

To perform prediction of the upcoming Solar Cycle 25, we use four solar cycles of synoptic magnetic field data from 1977 to 2019. Following the procedure described in Section~\ref{Observations}, we obtained estimates of the annual variation of the toroidal and poloidal fields. The magnetic field observations were calibrated to approximate the periodic model solution (Fig.~\ref{fig:MFCalibr}). In addition, the sunspot number, calculated  from the toroidal field measurements, was normalized to the sunspot number amplitude of Solar Cycle 24 to link the magnetograms and sunspot number data (Fig.~\ref{fig:MagSN}). Figure~\ref{fig:SC25c4predTorPol} shows the results of prediction for the toroidal (panel a) and poloidal (panel b) fields. The sunspot number forecast is shown in Figure~\ref{fig:SC25Pred4cyclesTPNS}. As expected, the forecast for the toroidal fields (and the sunspot number) is more accurate for the Northern hemisphere than for the Southern hemisphere because of smaller discrepancies between the model solution and observations at the end of Cycle 24.

The model solutions show strong variation in the toroidal fields near and after 2026.5 (red curves, Figs~\ref{fig:SC25c4predTorPol},~\ref{fig:SC25Pred4cyclesTPNS}a,b). These strong variations indicate that prediction uncertainties significantly increase after 2026.5. Thus, the sunspot number prediction for SC25 in the Northern hemisphere is about 30 (that is $\sim 50$\% weaker than SC24) with an error of 15 -- 20\% and about 25 for Southern hemisphere ($\sim 65$\% weaker than SC24) with error 25 -- 30\%. The solar maximum is expected during 2024 -- 2026 in the Northern hemisphere, and during 2024 -- 2025 in the Southern hemisphere. The total sunspot number predicted for the SC25 maximum is about 50 (for the v2.0 sunspot number definition) with an error estimate of $\sim 15-30$\%. It is important to keep in mind that in this case the total sunspot number forecast combines two separate predictions for each hemisphere. 

\section{Discussion and conclusions}\label{discussion}
Building accurate forecasts of solar activity requires knowledge of the non-linear multi-scale interactions of turbulent and large-scale flows and magnetic fields, which are described by global MHD models. At the present time, long-term synoptic observations provide evolution of magnetic fields and sunspots on the solar surface, while very limited information is available about long-term subsurface dynamics. In addition, the absence of realistic global models of the Sun requires finding a way to estimate solar activity evolution using simplified models. In such a state of limited knowledge of the global dynamics, shortage of observational data, and incomplete models, the data assimilation approach provides an efficient way to combine the data and models while taking into account uncertainties of both the models and observations. Currently, data assimilation includes a wide range of mathematical methods applicable to diverse problems in weather prediction, planetary sciences, fluid dynamics, etc. \citep[e.g.][]{Evensen1997,Kalnay2002}.  

Previously, we applied the Ensemble Kalman Filter method \citep{Evensen1997} to predict SC24 by using the sunspot data time-series and a simplified Parker-Kleeorin-Ruzmaikin \citep[PKR, ][]{Parker1955,Kleeorin1982} mean-field dynamo model \citep{Kitiashvili2009}. This approach allowed us to create a reliable forecast for the whole activity cycle \citep{Kitiashvili2008a}.  However, it is important to develop capabilities for predicting not only the general properties of solar activity in terms of the sunspot number, but also to forecast the evolution of solar magnetic fields including hemispheric activity. In this case, we have to deal with relatively short series of available observational data.

Recent studies based on the sunspot number data series \citep{Kitiashvili2019} show the possibility of obtaining a reasonable solar cycle prediction by assimilating the data for only two previous solar cycles. This encouraged us to apply the previously developed methodology to synoptic magnetogram data, available for four solar cycles. For this work, we combined all available synoptic observations from the National Solar Observatory and the SOHO and SDO space mission archives. The data were used to estimate hemispheric variations of the toroidal and poloidal field components and assimilate them into the dynamo model for prediction of the hemispheric solar activity. The dynamo model includes an additional parameter, the global magnetic helicity, for which there is no observational data, and which is recovered in the data assimilation procedure.
 
First, we performed three test predictions of SC23 and SC24 using different numbers of the available cycles (two and three). The goal of these tests was to evaluate the predictive capabilities of the method. For instance, the test prediction of SC23 was done using the synoptic magnetograms only for the period 1977.5 to 1996.5. However, it is found that an accurate prediction can be made only by including the data for 1975.5 and 1976.5 corresponding to the solar minimum period. This testing was performed by adding synthetic data for this period. As we have known from the previous studies \citep{Kitiashvili2008a,Kitiashvili2019}, the discrepancy between the model initial conditions and the observations is less important for assimilation of longer time series. In this case, it is critical to have a close phase match between the observed variations and the model solution. 

\begin{figure}[h]
	\begin{center}
		\includegraphics[scale=0.9]{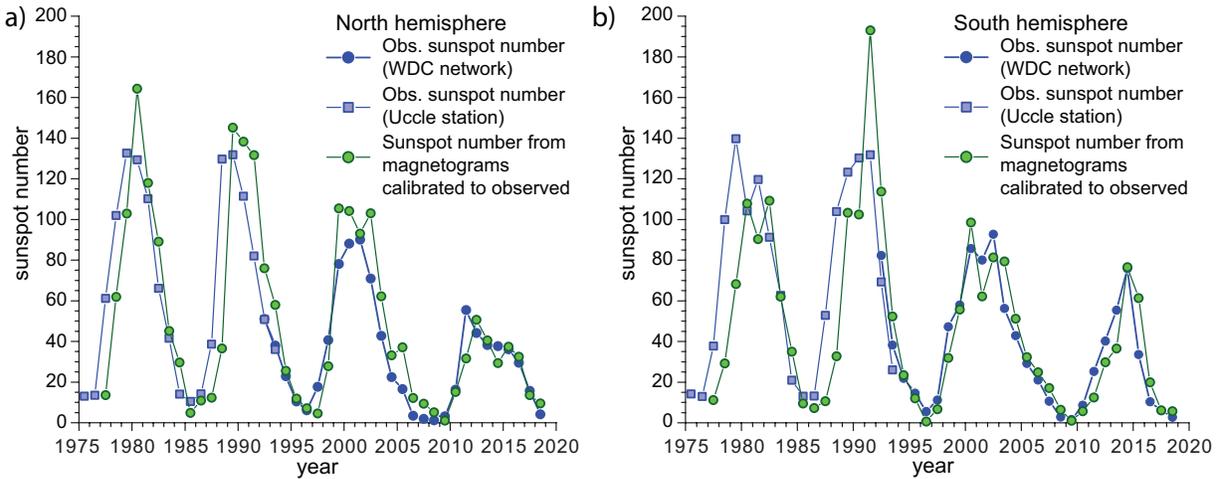}
	\end{center}
	\caption{Comparison of the observed annual sunspot number times-series from the WDC-SILSO network (blue circles, http://www.sidc.be/silso/datafiles), and Uccle station (blue rectangles) with the calibrated annual sunspot number times-series estimated from the synoptic magnetogramms (green circles) for: a) the Northern hemisphere and b) Southern hemisphere. \label{fig:MagSN}}
\end{figure}

\begin{figure}[h] 
	\begin{center}
		\includegraphics[scale=0.9]{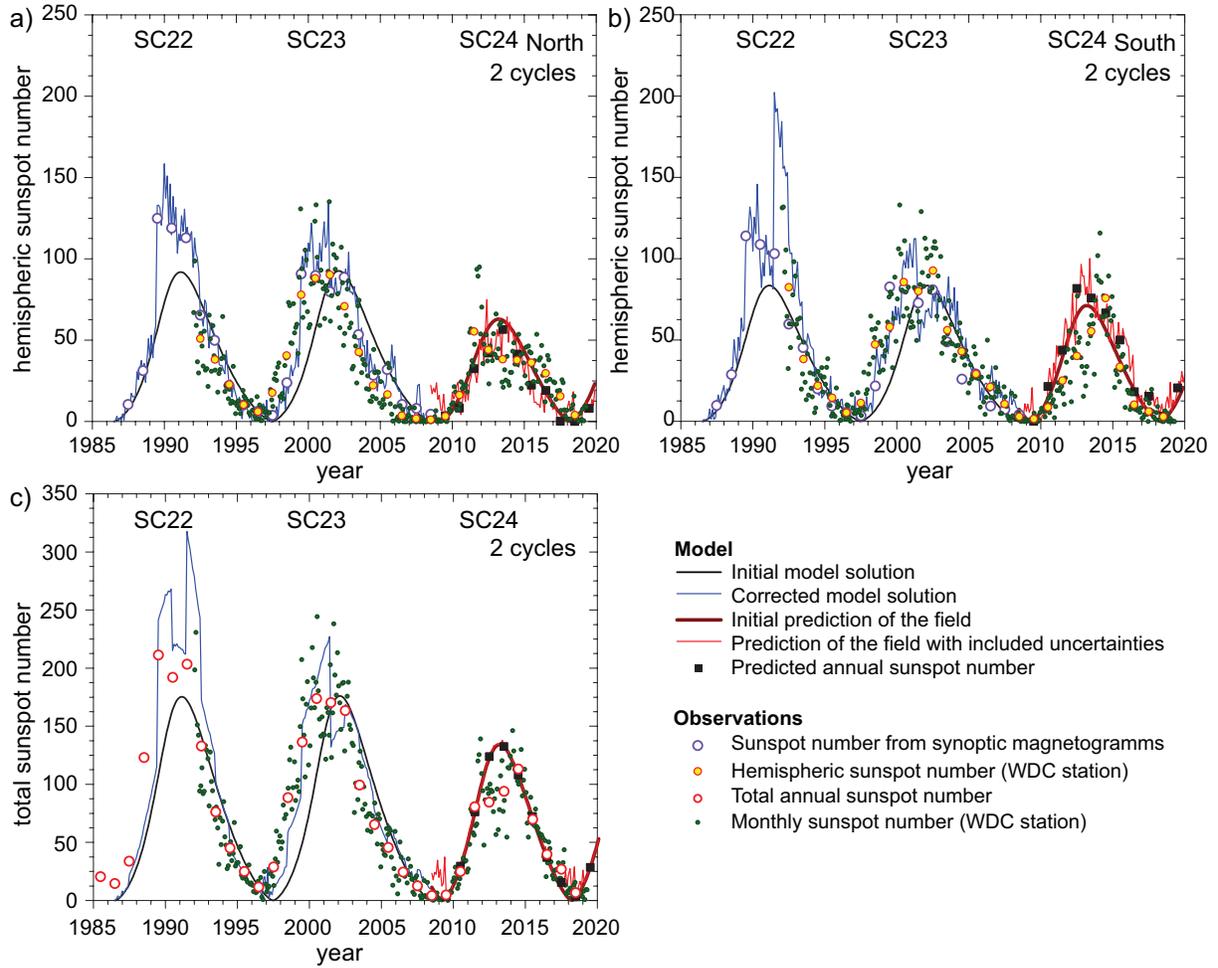}
	\end{center}
	\caption{Prediction of the sunspot number variations during SC23 for the Northern (panel a), Southern hemisphere (b), and total sunspot number (panel c) assuming availability of observations only for two previous cycles (case 1). \label{fig:SC24c2SN}}
\end{figure}

It is important to note that despite a good qualitative correlation between the mean toroidal field and the sunspot number, the exact quantitative relationship is not known. We used the three-halves law, $SSN \sim B_t^{3/2}$ \citep{Bracewell1988}, but this relation leads to some noticeable deviations (Fig.~\ref{fig:MagSN}) that introduce additional uncertainties in prediction of the sunspot number cycle. In addition, the total sunspot number variations are obtained by combining results for each hemisphere, which already include some uncertainties. It sometimes causes cancellation of the uncertainties and thus improves the forecast, but sometimes amplifies these uncertainties. For instance, the sunspot number predicted for SC23 in the Southern hemisphere is greater than the actual value, but the predicted total sunspot number matches the observations quite well. Contrarily, in the case of SC24 prediction based on 2 cycles, the sunspot number forecast for both hemispheres was in agreement with observations, but the total sunspot number variations were overestimated (Fig.~\ref{fig:SC24c2SN}).

To summarize, we can identify the following primary results of the data assimilation approach:
	\begin{itemize}
	\item Using two cycles of the synoptic magnetograms can provide a reasonable forecast of the solar activity for the following solar cycle.
	\item Including additional observations improves the predictive capability.
	\item Taking into account poloidal field observations can noticeably improve the forecast, particularly in the case when the data of three preceding cycles are assimilated in the model.
	\item Forecasted hemispheric toroidal field variations are in good agreement with observations, at least up to the following solar maximum, and often make a reasonable prediction for the whole activity cycle.
	\item Forecasted poloidal fields are in good agreement with observations for up to two years in the case of assimilation of data for two preceding activity cycles, and for about three years if data for three cycles is assimilated.
	\item According to the presented analysis, the next Solar Cycle 25 will be weaker than the current cycle and will start after an extended solar minimum during 2019 -- 2021. The maximum of activity will occur in 2024 -- 2025 with a sunspot number at the maximum of about $50\pm 15$ (for the v2.0 sunspot number series) with an error estimate of ~30\%.
	\item SC25 will start in the Southern hemisphere in 2020 and reach maximum in 2024 with a sunspot number of $\sim 28$ ($\pm 10$\%). Solar activity in the Northern hemisphere will be delayed for about 1 year (with error of $\pm 0.5$~year) and reach maximum in 2025 with a sunspot number of $\sim 23 \pm 5$ ($\pm 21$\%). 
\end{itemize}

The presented results encourage future development of the data assimilation methodology for more detailed dynamo models and more complete data sets.

{\bf Acknowledgment.} The work is supported by NSF grant AGS-1622341.
\newpage

\begin{table}
\begin{center}
	\begin{tabular}{| c | c | c | c |}
		\hline
		{\bf predicted} & {\bf \# cycles} & {\bf North} & {\bf South} \\
		{\bf cycle} & {\bf used} & {\bf Hemisphere} & {\bf Hemisphere} \\
		\hline
		23 & 2 & $3.96$\% & $0.75$\% \\
		24 & 2 & $0.86$\% & $-1.59$\% \\
		24 & 3 & $-8.6$\% & $10.5$\% \\	
		\hline
	\end{tabular}
\end{center}
\caption {Differences in the predicted sunspot number for SC23 and SC24 between the cases when both the toroidal and poloidal field components and only the toroidal field component were assimilated. \label{table}}
\end{table}

\newpage

\end{document}